\documentclass[a4paper, 10pt, aps, showkeys, showpacs, nofootinbib, twocolumn]{revtex4-2}
\usepackage{graphicx}
\usepackage{amsfonts}
\usepackage{amssymb}
\usepackage{amsbsy}
\usepackage{amsmath}
\usepackage{mathrsfs}
\usepackage{latexsym}
\usepackage{bm}
\usepackage{subfigure} 
\usepackage{wasysym}
\usepackage{mathbbol}
\usepackage{bigints} 
\usepackage{fontawesome}
\usepackage{natbib}
\usepackage{xcolor}
\usepackage{footnote}
\usepackage[colorlinks = true, linkcolor=blue, citecolor=blue, urlcolor=cyan]{hyperref}
\usepackage{orcidlink}

\begin{document}

\title[Thermal properties of accretion flows]{Accretion flow in deformed Kerr spacetime: Spectral energy distributions from free-free emission}

\author{Subhankar Patra \orcidlink{0000-0001-7603-3923}}
\email{psubhankar@iitg.ac.in}

\author{Bibhas Ranjan Majhi \orcidlink{0000-0001-8621-1324}}
\email{bibhas.majhi@iitg.ac.in}

\author{Santabrata Das \orcidlink{0000-0003-4399-5047}}
\email{sbdas@iitg.ac.in}

\affiliation{Department of Physics, Indian Institute of Technology Guwahati, Guwahati 781039, Assam, India}

\date{\today}

\begin{abstract}
	In this paper, we study the properties of accretion flow including its spectral features in Johannsen and Psaltis (JP) non-Kerr spacetime. In doing so, we numerically solve the governing equations that describe the flow motion around the compact objects in a general relativistic framework, where spin ($a_{k}$) and deformation parameters ($\varepsilon$) demonstrate the nature of the central source, namely black hole (BH) or naked singularity (NS). With this, we obtain all possible classes of global accretion solutions ($i. e.$, O, A, W and I-type) by varying the energy ($E$) and angular momentum ($\lambda$) of the relativistic accretion flow, and examine the role of thermal bremsstrahlung emission in studying the spectral energy distributions (SEDs) of the accretion disc. We divide the parameter space in $\lambda-E$ plane in terms of the different classes of accretion solutions for BH and NS models. We further calculate the disc luminosity ($L$) corresponding to these accretion solutions, and observe that I-type solutions yield higher $L$ and SEDs than the remaining types of solutions for both BH and NS models. For BH model, SEDs for W and I-type solutions differ significantly from the results for O and A-type solutions for low $E$ values. On the contrary, for NS model, SEDs for different accretion solutions are identical in the whole parameter space of $\lambda$ and $E$. We also examine the effect of $\varepsilon$ on the SEDs and observe that a non-Kerr BH yields higher SEDs than the usual Kerr BH. Finally, for accretion solutions of identical $E$ and $\lambda$, we compare the SEDs obtained from BH and NS models, and find that naked singularity objects produce more luminous power spectra than the black holes.                     	 	 
\end{abstract}




\maketitle
  
\section{Introduction}
Active galactic nuclei (AGNs) and black hole X-ray binaries (BHXRBs) are highly energetic sources in the Universe. These sources emit radiation due to the accretion of matter onto the central black holes \cite{pringle-1981, frank-2002, netzer-2013, abramowicz-2013}. Transition between different spectral states (Low/Hard, High/Soft, intermediate) of the emitted radiation are identified through the spectral properties of the accretion disc \cite{esin-1998}. For example, in the “Low/Hard” (LH) state, blackbody components are negligible ($i.e.$, low disc luminosity) and non-thermal components maximize the power in the high energy side. However, for the “High/Soft” (HS) state, exceptionally high luminous disc yields the multi-temperature blackbody components and the power-law components maximize power in the low energy side. In between LH and HS states, intermediate (IM) states also do exist in the disc spectrum. Indeed, a detailed spectral analysis of the accretion disc help to find black hole parameters ($i.e.$, mass and spin) \cite{aschenbach-2007, gou-2009, molla-2016a, molla-2016b, nandi-2018, das-2021, mondal-2022a, mondal-2022b, palit-2023, mondal-2023, heiland-2023} and disc properties (e.g., mass accretion rate, Comptonizing corona size, inner disc radius, inclination angle, quasi periodic oscillation (QPO) frequency, and photon index, etc.) \cite{nandi-2018, sreehari-2020, das-2021, Sriram-2021, majumder-2022, mondal-2022a, mondal-2022b, palit-2023, mondal-2023, heiland-2023, dhaka-2023, Rawat-2023a, Rawat-2023c} as well.           

After gravitational wave detection in LIGO and Virgo
collaboration \cite{abbott-2016a, abbott-2016-b} and the observation of black hole images of supermassive black holes by the event horizon telescope collaboration \cite{akiyama-2019-L1, akiyama-2019-L2, akiyama-2019-L3, akiyama-2019-L4, akiyama-2019-L5, akiyama-2019-L6}, the essence of theoretical astrophysics is increased a lot to understand distinctive observational features from the usual Kerr black holes. In the most straightforward Einstein’s theory of general relativity, several fundamental problems arise regarding the nature of singularities, dark energy, dark matter and quantization of gravitational interactions, etc. to address these issues. In these circumstances, theoretical research groups invent the modified gravity theory \cite{Sotiriou-2008, DeFelice-2010, Sullivan-2019}, higher curvature \cite{Deser-2002} and brane-world gravity \cite{Maartens-2010}, etc. Nowadays, there is a large number of studies have been done on the alternative gravity theory \cite{Bambi-2011-a, Bambi-2011-b, Bambi-2011-c, Liu-2012, Chen-2012, Bambi-2012, Krawczynski-2012}. These gravity theories do not render the no-hair theorem false \cite{Robinson-1975, Carter-1971}. Moreover, the spacetime outside a horizon-less compact object is characterized by the naked singularity or close timelike loops, found to be regular \cite{atamurotov-2013}. One such emerging alternative gravity theory is the Johannsen and Psaltis (JP) spacetime \cite{johannsen-2011}. This metric is described by a deformation parameter ($\varepsilon$), in addition to the mass ($M_{\rm BH}$) and spin ($a_{k}$) of the black hole \cite{atamurotov-2013}. In this spacetime, the central object can become naked singularity depending on the spacetime parameters ($a_{k}, \varepsilon$)~\cite{Bambhaniya-2021}. In the last decade, theorists have largely focused on the JP non-Kerr spacetime and analyzed different strong gravity signatures~\cite{Bambi-2011-d, Rezzolla-2014, Yagi-2016, Younsi-2016, Chen-2022, Patra-2022}.

Investigation of the nature of central singularity in the strong gravitational field is fascinating through the modern day’s highly precise observational data. The distinction between the physical properties of the accretion disc around the black hole and naked singularity has been reported in~\cite{Chowdhury-2012, Joshi-2020, Tahelyani-2022}. However, these studies do not include transonic accretion flows~\cite{abramowicz-1981, liang-1980}, which is yet to be explored. Motivated from this, we perform the present work, where thermal properties of the accretion disc in the JP non-Kerr spacetime has been studied. In this analysis, we consider the thermal bremsstrahlung emission from the hot accreting plasma around the supermassive compact objects. Even though the theory of thermal bremsstrahlung is well established, an accurate expression of the emission coefficient and its applicability in different temperature regimes is still an unresolved problem. In these circumstances, several formulas have been developed in the literature (e.g., see \cite{Yarza-2020} and references therein). As the electron temperature is in the relativistic regime for hot accretion flow (HAF), we use an approximate analytical expression of thermal bremsstrahlung emissivity derived by Novikov and Thorne~\cite{novikov-1973}. We find that the relativistic bremsstrahlung emissivity overestimates the disc luminosity compared to the non-relativistic model (NR-model). Therefore, the relativistic model (R-model) for thermal bremsstrahlung is used in our analysis.

Motivating with this, in this work, we obtain the transonic accretion solutions under the frameworks of general relativistic hydrodynamics and relativistic equation of state (REoS). We calculate the disc luminosity and spectral energy distributions (SEDs) corresponding to different flow topologies (O, A, W and I-types) in the black hole spacetime. We infer that I-type solutions produce high luminous power spectra compared to O, A and W-type solutions. Moreover, SEDs for W and I-type solutions significantly differ from O and A-type solutions, particularly for low flow energy ($E$). Also, the effect of deformation parameter ($\varepsilon$) on the luminosity spectrum has been investigated. We notice that SEDs increase with $\varepsilon$, indicating that an accretion disc around a non-Kerr black hole generates more luminous power spectrum than the usual Kerr black hole. We further extend our analysis for a naked singularity object embedded in the JP non-Kerr metric. Like the black hole model, we notice that SEDs for I-type solutions are greater than the other flow topologies. But, in contrast to the black hole model, luminosity distributions for W and I-type solutions moderately differ from O and A-type solutions. 

Most importantly, from a comparative study between the SEDs, we wish to emphasize that a naked singular object can produce a high luminous power spectrum at both low and high-frequency regions compared to the black hole. These results open up a window to isolate black holes and naked singularity objects through the spectral analysis of accretion disc.                            

This paper is arranged as follows. In Section~\ref{sec:flow-hydrodynamics}, we construct the relativistic hydrodynamic equations of a perfect fluid in the JP non-Kerr spacetime. In Section~\ref{sec:bremsstrahlung}, we develop the mathematical framework for thermal bremsstrahlung and also depict the importance of relativistic model over the non-relativistic model. Section~\ref{sec:SED-Kerr-BH} has investigated the luminosity distribution for different flow topologies in black hole spacetime. In Section~\ref{sec:deformed-spacetime}, we study the effect of spacetime deformation on the disc spectrum. A comparative study between the spectral energy distributions (SEDs) for black hole and naked singularity models has been studied in Section~\ref{sec:SED-BH-NS}. Finally, we summarize our findings in Section~\ref{sec:Summary}. 

\section{Flow Hydrodynamics}   
\label{sec:flow-hydrodynamics} 

In this paper, we carry out the analysis of thermal bremsstrahlung emission from accretion flow in a stationary and axisymmetric non-Kerr spacetime. In doing so, we study the governing flow equations using general relativistic hydrodynamics framework and relativistic equation of state (REoS). We consider a steady, inviscid and axisymmetric accretion flow.  

We adopt a unit system as $G = c = M_{\rm BH} = 1$, where $G$ is the gravitational constant, $c$ is the speed of light, and $M_{\rm BH}$ is the mass of the central source. In this unit system, Johannsen-Psaltis (JP) metric is expressed as~\cite{johannsen-2011},
\begin{equation}
\begin{split}
\label{eq:deformed-Kerr-metric}
ds^{2} & = -~(1-\frac{2M_{\rm BH}r}{\Sigma})(1+\mathfrak{h}(r,\theta))dt^{2}
\\& - \frac{4M_{\rm BH}r a_{ k}\sin^{2}\theta}{\Sigma}(1+\mathfrak{h}(r,\theta))dtd\phi
\\& + \frac{\Sigma(1+\mathfrak{h}(r,\theta))}{\Delta + a_{ k}^{2}\mathfrak{h}(r,\theta) \sin^{2}\theta} dr^{2} + \Sigma d\theta^{2} 
\\& + \left[\Sigma + a_{ k}^{2}(1+\mathfrak{h}(r,\theta))(1+\frac{2M_{\rm BH}r}{\Sigma})\sin^{2}\theta\right]\sin^{2}\theta d\phi^{2},
\end{split}
\end{equation}
where $a_{k}$ is the spin parameter, $\varepsilon$ is the deformation parameter, $\Sigma = r^{2} + a_{k}^{2}\cos^{2}\theta$ and $\Delta = r^{2} - 2M_{\rm BH}r + a_{k}^{2}$. Here, $\mathfrak{h}(r,\theta) ~(= \varepsilon M_{\rm BH}^{3}r/\Sigma^{2})$ accounts parametric deviation to the usual Kerr solution. For $\varepsilon = 0$, the metric in Eq.~(\ref{eq:deformed-Kerr-metric}) reduces into the original Kerr metric~\cite{boyer-1967}.

Here, we briefly mention about the origin of the JP metric and its importance in explaining various physical phenomena. This will lay down the basis of considering the JP metric in the context of accretion dynamics. LIGO and Virgo measurements on both electromagnetic and gravitational wave spectra reveal several strong gravity signatures, namely dark energy, dark matter, nature of singularities, quantization of gravitational waves, close time like curves, etc. Understanding these properties within the Einstein's gravity framework faces difficulties. The other gravity theories (commonly known as modified theories) like $f(R)$, Gauss-Bonnet gravity, inclusion of higher curvature, brane-world gravity, etc. play significant role in illuminating the issues. In these gravity theories, one always finds exact metric solutions by solving a sets of field equations. These theories mainly contain modifications of the general relativity (GR). Alternative approach to examine the possible explanations of the above mentioned properties is to find a spacetime by perturbing the seed solutions of GR. The perturbation is carried out mainly based on the underlying boundary conditions and the parameter of the perturbation is decided by means of the observations. Note that in the latter approach, the metric is not obtained by perturbing the seed action of the theory and hence, one does not know the actual gravitational action. In a nut-shell, there are numerous prescriptions available in the literature for finding a fruitful theoretical model. However, none of the above approaches has complete dominance over the other till date. Accordingly, it still remains open to adopt any of these approaches while investigating the physical phenomena. JP metric is one such example under the perturbative method. In this case, the usual Kerr solution is perturb and the perturbation parameter (namely the deformation parameter $\varepsilon$) is constrained from the observation. According to the no-hair theorem, any parametric deviation to the Kerr metric is no longer the solution of Einstein's gravity or modified gravity~\cite{johannsen-2011, Konoplya-2016}. Therefore, as the JP metric includes a deformation parameter (upper limit to this parameter is bounded through the observation~\cite{atamurotov-2013}), it must belong to some other gravity theory called alternative gravity, rather than Einstein or modified gravity theory. Interestingly, we do not have any such definite gravity theory that invoked the JP metric. Since this metric explains several novel features mentioned above, we hope it will also affect fluid dynamics in the accretion disc. One such important outcome of this metric is the naked singular solution, which is characterized by the close timelike orbits outside the singularity with negative precession~\cite{Bambhaniya-2021}. As our goal is to explore the accretion processes around both black holes and naked singularity exotic objects, we choose the JP metric in our present study. Of course one can consider the exact solutions of the modified theories of gravity to investigate accretion properties.

In the accretion disc model, we consider the motion of
a perfect fluid along the equatorial plane ($i.e.$, $\theta = \pi/2$) of central object. Under this assumption, the time translation and azimuthal symmetries in the metric in Eq.~(\ref{eq:deformed-Kerr-metric}) provide two conserved quantities associated with the flow as $E = - (e + p)u_{t}/\rho$ and $\mathcal{L} = (e + p)u_{\phi}/\rho$, where $E$ and $\mathcal{L}$ are refer the Bernoulli function and bulk angular momentum per unit mass, respectively. Here, $e$ is the total internal energy density, $p$ is the pressure, $\rho$ is the mass-density, $u_t$ and $u_\phi$ are $t$ and $\phi$ components of the covariant four-velocity ($u_{k}$), respectively. Therefore, the ratio $\lambda = E/\mathcal{L} = - u_\phi/u_t$, defined as the specific angular momentum, is also a conserved quantity along the stream lines of the flow. In this framework, the radial-momentum equation of the flow in a co-rotating frame ($i.e.$, a frame rotate with the flow angular velocity $\Omega = u^\phi/u^t$) is given by~\cite{dihingia-2018},
\begin{equation}
\label{eq:radial-momentum-equation}
\gamma_{v}^{2}v\frac{dv}{dr} + \frac{1}{e + p}\frac{dp}{dr} + \frac{d\Phi^{\rm eff}}{dr} = 0,
\end{equation}
where $\Phi^{\rm eff}$ is the effective potential of the system and $\gamma_{v} = 1/\sqrt{1 - v^2}$ is the Lorentz-factor corresponding to the radial velocity $v = \sqrt{u^{r}u_{r}/(u^{t}u_{t}(\Omega \lambda - 1))}$~\cite{lu-1985}. An explicit form of $\Phi^{\rm eff}$ is obtained as,
\begin{equation}
\begin{split}
\label{eq:Phi}
\Phi^{\rm eff} & = \frac{1}{2}\ln\left[\frac{(\Delta+a_{\rm k}^{2}\mathfrak{h})(1+\mathfrak{h})r}{Q}\right];~Q = a_{\rm k}^{2}(r+2)(1+\mathfrak{h})
\\& - 4a_{\rm k}\lambda(1+\mathfrak{h}) -\lambda^{2}(r-2)(1+\mathfrak{h}) + r^{3}.
\end{split}	
\end{equation}

The energy generation equation (first law of thermodynamics) is given by~\cite{dihingia-2018},
\begin{equation}
\label{eq:energy-generation}
\frac{e + p}{\rho}\frac{d\rho}{dr} - \frac{de}{dr} = 0.
\end{equation}

From the conservation of mass flux ($i.e.$, $\nabla_k(\rho u^k) = 0$), we calculate the mass accretion rate of the flow as,
\begin{equation}
\label{eq:mass_accretion}
\dot{M} = 4\pi \rho v \gamma_{v} H\sqrt{(\Delta + a_{\rm k}^{2}\mathfrak{h})(1+\mathfrak{h})},
\end{equation} 
where $H$ is the local half thickness of accretion disc. 

Considering hydrostatic equilibrium in the vertical direction of the disc, we obtain $H$ as~\cite{lasota-1994,  riffert-1995, Peitz-1996},
\begin{equation}
\label{eq:hafl_thickness}
H = \sqrt{\frac{pr^{3}}{\rho F}};~F = \frac{(r^{2} + a_{\rm k}^{2})^{2} + 2\Delta a_{\rm k}^{2}}{(1 - \Omega\lambda)\left((r^{2} + a_{\rm k}^{2})^{2} - 2\Delta a_{\rm k}^{2}\right)}.
\end{equation}

Since the flow temperature vary in a wide range within the disc, we adopt a relativistic equation of state (REoS) with variable adiabatic index $\Gamma$~\cite{chattopadhyay-2009}. Integrating Eq.~(\ref{eq:energy-generation}) by using REoS, we get the entropy accretion rate ($\mathcal{\dot{M}}$) of flow as~\cite{chattopadhyay-2016, kumar-2017},
\begin{equation}
\label{eq:entropy_accretion_rate}
\begin{split}
\mathcal{\dot{M}} = \frac{\dot{ M}}{4\pi\mathcal{K}} & = v\gamma_{v}H\sqrt{(\Delta+a_{k}^{2}\mathfrak{h})(1+\mathfrak{h})}
\\& \times \Theta^{3/2}(3\Theta + 2)^{3/4}(3\Theta+2m_{p}/m_{e})^{3/4}\exp{(\chi)},
\end{split}
\end{equation}  
where $\mathcal{K}$ is the entropy constant, $\Theta = k_{B}T/(m_{e}c^{2})$ is the dimensionless temperature, $k_B$ is the Boltzmann constant and $T$ is the flow temperature in Kelvin. Here, $\chi = (f - 1 - m_{p}/m_{e})/(2\Theta)$ and the quantity $f$ is expressed in terms of $\Theta$ as~\cite{chattopadhyay-2009},
\begin{equation}
f = 1 + \frac{m_p}{m_e} + \Theta\left[\left(\frac{9\Theta + 3}{3\Theta + 2}\right) + \left(\frac{9\Theta + 3m_p/m_e}{3\Theta + 2m_p/m_e}\right)\right],
\end{equation}
where $m_{e}$ and $m_p$ are the mass of electron and proton, respectively.

Now, it should be emphasized that accretion solutions can possess multiple critical points ($i.e.$, $r_{\rm in}$ and $r_{\rm out}$) for a given set of input parameters ($a_{k}, \varepsilon, E, \lambda$). There are many studies in the literature that mainly explore two kinds of accretion solutions containing multiple critical points~(e.g., see~\cite{dihingia-2018, dihingia-2018b, dihingia-2020, dihingia-2020a} and references therein). In the first scenario, the solution passing through $r_{\rm in}$ fails to connect the outer edge ($r_{\rm edge}$) to the inner edge ($r_{0}$) of the disc (called close solution), however the solution containing $r_{\rm out}$ smoothly extended from $r_{\rm edge}$ to $r_{0}$ (called open solution). In the second scenario, the solution passing through $r_{\rm out}$ remains closed, while the same crossing $r_{\rm in}$ remains open. These two flow topologies are characteristically different, which is quantified by means of their entropy content. In this work, the accretion flow is adiabatic in nature and the entropy accretion rate ($\mathcal{\dot{M}}$) is expressed in Eq.~(\ref{eq:entropy_accretion_rate}) which remains constant along the flow streamline. In Section~\ref{sec:SED-Kerr-BH}, we elaborately discuss how the entropy content of the flow and other physical criteria classify various types of accretion solutions.
	
In the accretion dynamics, mass accretion rate ($\dot{ M}$) of the flow is usually taken as constant through out the disc. Therefore, using $d{\dot{ M}}/dr = 0$, we get the temperature gradient of flow as,
\begin{equation}
\label{eq:temperature-gradient}
\begin{split}
\frac{d\Theta}{dr} = - \frac{2\Theta}{f^{\prime} + 1} \left[  \frac{\gamma_{v}^{2}}{v}\frac{dv}{dr} + N_{11} + N_{12}\right],
\end{split}
\end{equation}
where  
\begin{equation}
\begin{split}
& f^{\prime} = \frac{df}{d\Theta};~N_{11} = \frac{5}{2r} + \frac{r-a_{\rm k}^{2}(1+\mathfrak{h})}{r(\Delta + a_{\rm k}^{2}\mathfrak{h})} - \frac{1}{2F}\frac{dF}{dr}; 
\\& 
N_{12} = - \frac{3\varepsilon}{2r^{4}}\left( \frac{a_{\rm k}^{2}}{\Delta + a_{\rm k}^{2}\mathfrak{h}} + \frac{1}{1+\mathfrak{h}}\right).
\end{split}
\end{equation}

Now, using Eqs.~(\ref{eq:radial-momentum-equation}) and (\ref{eq:temperature-gradient}), we obtain the wind equation involving sound speed $C_{s} = \sqrt{\Gamma p/(e + p)}$ as, 
\begin{equation}
\label{eq:velocity-gradient}
\frac{dv}{dr} = \frac{\mathcal N}{\mathcal D},
\end{equation} 
where
\begin{equation}
{\mathcal N} = \frac{2C_{s}^{2}(N_{11} + N_{12})}{\Gamma + 1} - \frac{d\Phi^\text{eff}}{dr};~{\mathcal D} = \gamma_{v}^{2}\left[v - \frac{2C_{s}^{2}}{(\Gamma + 1)v}\right].
\end{equation}

For transonic accretion solutions, flow must pass through at least one critical point ($r_c$), where $dv/dr$ takes the form $``0/0"$~\cite{abramowicz-1981, liang-1980}. Hence, at the critical point we apply l$'$H\^{o}pital's rule to calculate $(dv/dr)_{r_c}$.

We obtain the accretion solutions by solving Eq.~(\ref{eq:velocity-gradient}) and these solutions are very useful to calculate the bremsstrahlung emission spectrum in a non-Kerr spacetime. However, when we set the deformation $\mathfrak{h} = 0$ ($i.e.$, $\varepsilon = 0$), we get back to the usual flow hydrodynamics in the Kerr spacetime~\cite{dihingia-2018}. The detail discussion of the thermal bremsstrahlung emissivity from a system of hot accreting plasma has been discussed in the next section.  

\section{Bremsstrahlung}
\label{sec:bremsstrahlung}

In our analysis, we consider the bremsstrahlung (free-free) emission from the accelerating electrons in the field of heavy ions. For simplicity, we consider negligible deviation of the electron's path during a close encounter with an ion. Further, we consider Maxwell-Boltzmann (MB) speed distribution of the thermal electrons. Under these assumptions, we obtain an expression of the non-relativistic bremsstrahlung emissivity (power density) at a particular frequency $\nu$ as \cite[]{rybicki-1991},
\begin{equation}
\begin{split}
\mathcal{E}_{\nu}^{\rm ff, ei} = \frac{32\pi e^{6}}{3m_{e}c^{3}}\sqrt{\frac{2\pi}{3m_{e}k_{\rm B}}}n_{e}n_{i}Z^{2}T_{e}^{-1/2}e^{-h\nu/k_{\rm B}T_{e}}\bar{g}_{\rm ff}^{\rm ei},
\end{split}
\label{eq:NR-br-emissivity}
\end{equation}      
where $T_{e}$ is electron temperature, $n_{e}$ is electron number density, $n_{i}$ is ion number density, $Z$ is ion atomic number and $h$ is the Planck constant. The thermally averaged electron-ion Gaunt factor $\bar{g}_{\rm ff}^{\rm ei}$ includes quantum mechanical correction to the classical electrodynamics. Depending on the energy of emitting electrons, $\bar{g}_{\rm ff}^{\rm ei}$ varies from $1$ to $5$ \cite{karzas-1961, brussaard-1962}. However, we take $\bar{g}_{\rm ff}^{\rm ei} = 1.2$~\cite{Yarza-2020} throughout our analysis. Moreover, we set $Z = 1$ for hydrogen plasma.

When the plasma temperature is in the non-relativistic regime ($k_{\rm B}T_{e} < m_{e}c^{2}$), radiations due to the encounter between like particles e.g., electron-electron and ion-ion are negligible in comparison with the electron-ion emission. This is simply due to the conservation of dipole moment, and the contributions from quadrupole and higher order moments have very less intensity in comparison to the dipole part \cite{rybicki-1991}. But, when $k_{\rm B}T_{e} \geq m_{e}c^{2}$, electron-electron bremsstrahlung dominates the power over the electron-ion emission \cite[]{Svensson-1982, nozawa-2009, Yarza-2020}. In case of hot accretion flow (HAF), plasma temperature can vary from $10^9~\rm K$  to $10^{12}~\rm K$ \cite{Yuan-2014, dihingia-2020, das-2021, Sarkar-2022}. So, in this temperature range, relativistic effect and electron-electron emission can modify the emissivity in Eq.~(\ref{eq:NR-br-emissivity}). An approximate expression of the total bremsstrahlung emissivity is given by~\cite{novikov-1973},
\begin{equation}
\begin{split}
\mathcal{E}_{\nu}^{\rm ff} & = \frac{32\pi e^{6}}{3m_{e}c^{3}}\sqrt{\frac{2\pi}{3m_{e}k_{B}}}n_{e}n_{i}Z^{2}T_{e}^{-1/2}
\\& \times (1 + 4.4\times 10^{-10}T_{e})e^{-h\nu/k_{\rm B}T_{e}}\bar{g}_{\rm B},
\end{split}
\label{eq:R-br-emissivity}
\end{equation}
where $\bar{g}_{\rm B}$ is the thermally-averaged Gaunt factor, which is taken as $1.2$~\cite{Yarza-2020}. The additional terms in Eq.~(\ref{eq:R-br-emissivity}) incorporate the relativistic effect and electron-electron bremsstrahlung as well.

In general, the radiations from the hot plasma flowing through the disk come out isotropically. For an observer at static infinity, such radiation gets red-shifted while it travels into the immense potential well of the central compact object. In addition, emitting gas proceeds toward the observer or away from the observer due to disc rotation, producing Doppler shift. Therefore, resulting redshift factor ($1 + z$) depends on the combined effect of both the gravitational and Doppler part. For simplicity, we neglect the light-bending effect to the emitting photons. As a result, the redshift factor, defined as the ratio of emitted frequency $(\nu_e)$ to observed frequency $(\nu_{o})$, is calculated as~\cite{luminet-1979},
\begin{equation}
\begin{split}
\frac{\nu_e}{\nu_o} = 1 + z =  u^{t}\left(1 + \frac{r\Omega}{c}\sin{\theta_{0}}\sin{\phi}\right),
\end{split}
\label{eq:redshift}
\end{equation}
where $\theta_{0}$ is the inclination angle of the accretion disc with respect to the distant observer direction, which is taken as $45^{\circ}$ \cite{dihingia-2020a, Sen-2022}. To calculated $u^{t}$, one has to use the standard time-like condition $u^{k}u_{k} = -1$. After expanding the above condition, we normalize the equation by $-u^{t}u_{t}$. Following \cite{lu-1985, Peitz-1996}, we make use of the definition of azimuthal and radial components of three-velocity in a co-rotating frame as $v_{\phi} = \sqrt{u^{\phi}u_{\phi}/(-u^{t}u_{t})}$ and $v = \gamma_{\phi}\sqrt{u^{r}u_{r}/(-u^{t}u_{t})}$, respectively. Here, we consider positive $u^{t}$ just to ensure that flow energy ($E = -(e + p)u_{t}/\rho$) remains positive all throughout the flow. After mathematical simplification, we finally obtain $u^{t}$ in terms of input parameters as,
	\begin{equation}
		\label{eq:u_upper_t}
		\begin{split}
			u^{t} = \gamma_{v}\gamma_{\phi}\sqrt{\frac{r}{(2a_{k}\Omega + (r - 2))(1 + \mathfrak{h})}},	
		\end{split}
	\end{equation}
	where $\gamma_{\phi} ~[= 1/\sqrt{1-v_{\phi}^2} = 1/\sqrt{1 - \Omega\lambda}]$ and $\gamma_{v} ~[= 1/\sqrt{1-v^{2}}]$ are the Lorentz-factors corresponding to $v_\phi$ and $v$, respectively. Here, the angular velocity of the flow ($\Omega = u^{\phi}/u^{t}$) is given by,
	\begin{equation}
		\label{eq:Omega}
		\Omega = \frac{(2a_{k} + \lambda (r-2))(1+\mathfrak{h})}{r^{3} + (a_{k}^{2}(r+2) - 2a_{k}\lambda)(1+\mathfrak{h})}.
	\end{equation}

In the hot plasma, we consider the electron and ion number densities are identical. They relate with the mass density ($\rho$) as $n_{e} = n_{i} \approx \rho/m_{p}$. Therefore, the monochromatic disc luminosity for an observer at infinity is obtained from Eqs.~(\ref{eq:R-br-emissivity}) and (\ref{eq:redshift}) as,
\begin{equation}
\label{eq:monochromatic-L}
\begin{split}
L_{\nu_o} & = 2\int_{r_0}^{r_{\rm edge}}\int_{0}^{2\pi}\mathcal{E}_{\nu_{e}}^{\rm ff}Hr~dr d\phi
\\& = 4.875\times 10^{10}\bar{g}_{\rm B}~{\rm erg~s^{-1}~Hz^{-1}}  \int_{r_0}^{r_{\rm edge}}\int_{0}^{2\pi}\Big[\rho^{2}T_{e}^{-1/2}
\\& \times (1 + 4.4\times 10^{-10} T_{e})e^{-(1+z)h\nu_o/k_{\rm B}T_{e}}Hr\Big]dr d\phi, 
\end{split}
\end{equation}
where $r_0$ and $r_{\rm edge}$ are the inner and outer edges of the accretion disc, respectively.

Finally, we get the frequency-integrated disc luminosity by using Eq.~(\ref{eq:monochromatic-L}) as,
\begin{equation}
\begin{split}
\label{eq:bolometric-L}
L & = \int_{0}^{\infty}L_{\nu_o}d\nu_{o} 
\\& = 1.015\times 10^{21}\bar{g}_{\rm B}~{\rm erg~s^{-1}} 
\\& \times \int_{r_0}^{r_{\rm edge}}\int_{0}^{2\pi}\frac{\rho^{2}T_{e}^{1/2}(1 + 4.4\times 10^{-10} T_{e})Hr}{u^{t}\left(1 + \frac{r \Omega \sin{\phi}}{\sqrt{2}c}\right)}~drd\phi.
\end{split} 
\end{equation}

As the mass of electrons is less than the mass of protons, electron temperature ($T_{e}$) must be less than the proton temperature ($T_{p} = T$) at least in the vicinity of the central source. This findings are already addressed while studying the two temperature accretion flows \cite{shapiro-1976, colpi-1984, Chakrabarti-1995, rajesh-2010, Dihingia-2017b, dihingia-2018b, Sarkar-2018, Chael-2018, dihingia-2020, Sarkar-2022}. However, in this work, we take $T_{p}/T_{e} = T/T_{e} = 10$ throughout the accretion disc~\cite{Yarza-2020} for the purpose of simplicity. We consider a supermassive black hole of mass $M_{\rm BH} = 10^{6}M_{\odot}$, $M_{\odot}$ being the Solar mass. As the mass accretion rate ($\dot{M}$) of the flow is very low for supermassive black hole~\cite{Yuan-2014, Yarza-2020}, we set $\dot{M} = 10^{-5}\dot{M}_{\rm Edd}$, where $\dot{M}_{\rm Edd} = 1.39 \times 10^{18}M_{\rm BH}/M_{\odot}~\rm gm~s^{-1}$ is the  Eddington mass accretion rate. At this low $\dot{M}$, we neglect the effect of radiative cooling on the flow dynamics~\cite{ryan-2017, Yarza-2020}. Moreover, for the black hole (BH) model, inner and outer edges of the accretion disc are taken as $r_{0} = r_{\rm H}$ (event horizon) and $r_{\rm edge} = 1000$, respectively. 

In order to determine the horizon location, one needs to find the maximum root of the equation ${g^{rr}} (r=r_{\rm H})=0$. As the accretion flow is confined around the disk equatorial plane ($\theta=\pi/2$), we obtain the horizon ($r_{\rm H}$) by numerically solving the following equation,
	\begin{equation}
		\label{eq:event-horizon}
		r_{\rm H}^{5} - 2r_{\rm H}^{4} + a_{k}^{2}r_{\rm H}^{3} + a_{k}^{2}\varepsilon = 0.
	\end{equation}
	It is noteworthy that for naked singularity (NS) model, $r_{0}$ can not be determined apriori as NS does not contain a inner boundary similar to the horizon as in the case of BH. Hence, in this work, we estimate $r_{0}$ according to the model input parameters and choose $r_{\rm edge} = 1000$. A detailed discussion in obtaining $r_{0}$ for NS model is presented in Section~\ref{sec:SED-NS}.

\begin{figure}
	\centering
	\includegraphics[width=\columnwidth]{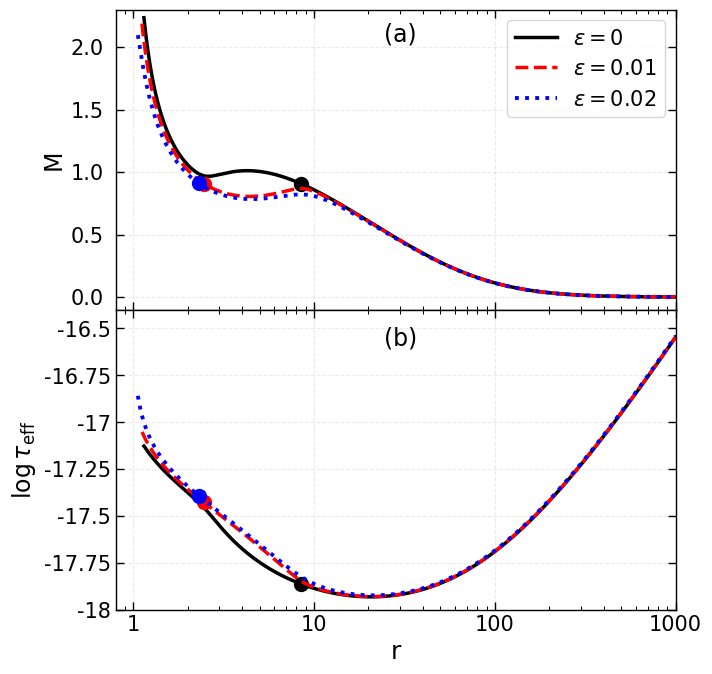}
	\caption{(a) Global accretion solutions ($M$ vs $r$ plots) for different deformation parameter ($\varepsilon$) and (b) plot of effective optical depth $(\tau_{\rm eff})$ as a function of radial distance ($r$) corresponding to the accretion solutions presented in the panel (a). In this figure, we choose $a_{k} = 0.99$ and $(\lambda, E) = (1.825, 1.0215)$. Here, critical points are marked by the filled circles. See the text for details.}  
	\label{fig:fig1}
\end{figure}

As the bremsstrahlung emission can only be detected when the emitting medium is optically thin, we need to calculate the optical depth of the accretion disc. We consider a photon generated deep inside the medium due to thermal emission. It suffers several coherent and isotropic scattering during its propagation into the medium and ends with a true absorption. One of the vital sources of opacity is the Thomson scattering of photons by the free electrons. The scattering optical depth is given by $\tau_{s} = \kappa_{s}\rho H$, where $\kappa_{s} \approx 0.4~\rm{cm^{2}~gm^{-1}}$ is the opacity coefficient. Here, the typical length scale of the medium is taken as the local half thickness of disc ($H$). Other significant opacity sources are bound-bound, bound-free (photoionization) and free-free absorption (inverse bremsstrahlung), which depend on the photon frequency, flow density, temperature and plasma composition. However, for a fully ionized medium, only free-free absorption is present \cite{maoz-2016}. The Rosseland mean opacity coefficient for free-free absorption is calculated from Eq.~(\ref{eq:R-br-emissivity}) as \cite{shapiro-2008}  
\begin{equation}
	\kappa_{\rm R}^{\rm ff} = 6.45\times 10^{22}\rho T_{e}^{-3.5}(1 + 4.4\times 10^{-10}T_{e})\bar{g}_{\rm R}~{\rm cm^{2}~gm^{-1}}, 
\end{equation}     
where, $\bar{g}_{\rm R}$ is the frequency-averaged Gaunt factor of $\bar{g}_{\rm B}$, which is of the order  unity. The corresponding absorption optical depth is given by $\tau_{a} = \kappa_{\rm R}^{\rm ff}\rho H$. Therefore, the effective optical depth of the medium is obtained as $\tau_{\rm eff} \approx \sqrt{\tau_{a}(\tau_{a} + \tau_{s})}$ \cite{rybicki-1991}.

Now, we intend to calculate the optical depth of the accretion disc for different deformation parameters ($\varepsilon$). In this case, input parameters are taken as $a_{k} = 0.99$ and $(\lambda, E) = (1.825, 1.0215)$. Flow topologies corresponding to $\varepsilon = 0, 0.01$ and $0.02$ are A, W, and I-types, respectively. In the subsequent sections, a clear prescription for isolating these flow topologies has been given. We calculate $\tau_{\rm eff}$ associated with the global solutions (see Fig.~\ref{fig:fig1}a) in the above flow topologies. Obtained results are plotted as a function of radial coordinate ($r$) in Fig.~\ref{fig:fig1}b. We see that the accretion disc remains effectively thin throughout the disc. Therefore, most photons leave the medium before being absorbed and fails to maintain thermal equilibrium with the matter.

\begin{figure*}
	\centering
	\includegraphics[width=\linewidth]{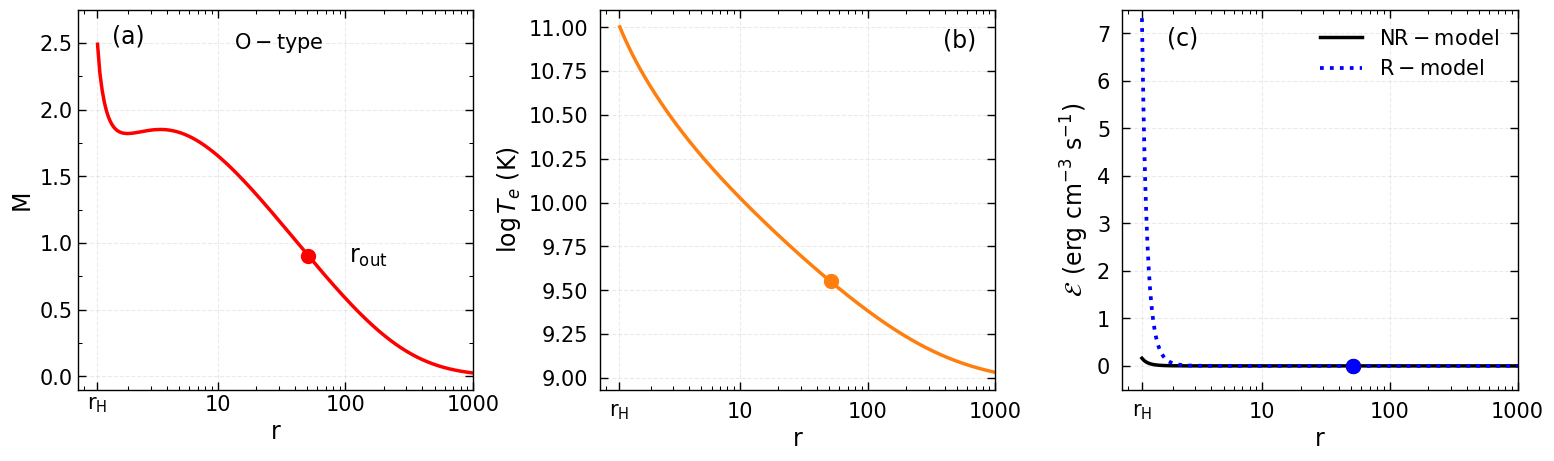}
	\caption{Plot of (a) Mach number ($M = v/C_{s}$), (b) Electron temperatures, and (c) frequency-integrated emissivity as function of radial distance ($r$). Here, we choose the input parameters as $a_{k} = 0.99$ and $\lambda = 1.85$ and $E= 1.005$. Critical point is marked using filled circle. See the text for details.} 
	\label{fig:fig2}
\end{figure*}

Next, we explore the relativistic effect and electron-electron emission on the emitted radiation in the Kerr spacetime. Here, we choose a rapidly-rotating Kerr black hole of spin $a_{k} = 0.99$. Further, angular momentum and energy of the flow are taken to be $\lambda = 1.85$ and $E = 1.005$, respectively. With these input parameters, we first obtain a flow solution ($i.e.$, Mach number ($M = v/C_{s}$) as function of radial distance ($r$)) by numerically solving the Eqs.~(\ref{eq:temperature-gradient}) and (\ref{eq:velocity-gradient}). We see that flow possesses outer critical point only at $r_{\rm out} = 51.2471$. The solution passes through $r_{\rm out}$ globally connects $r_{\rm H} = 1.1413$ and $r_{\rm edge} = 1000$, and often called it as O-type flow topology (see Fig.~\ref{fig:fig2}a). Temperature profile of the electrons ($T_{e}$ as a function of $r$) in the accretion disc is presented in Fig.~\ref{fig:fig2}b. We see that $T_e$ increases monotonically toward the disc's inner edge. After that the frequency-integrated emissivity ($\mathcal{E}$) is calculated using Eq.~(\ref{eq:NR-br-emissivity}) for non-relativistic (NR) model and using Eq.~(\ref{eq:R-br-emissivity}) for relativistic (R) model. Obtained results are plotted in Fig.~\ref{fig:fig2}c, where we present the variation of $\mathcal{E}$ as a function of $r$. In the figure, solid (black) and dotted (blue) curves represent $\mathcal{E}$ for NR and R-models, respectively. We find that for both cases, $\mathcal{E}$ does not differ too much from $r > 2$ to $r_{\rm edge}$. However, for $r_{\rm H} < r < 2$, the difference in $\mathcal{E}$ substantially increases as we proceed toward $r_{\rm H}$. Needless to mention that the above range in $r$ is not universal, which depends on the input parameters. But, we always get a significant difference in $\mathcal{E}$ very close to $r_{\rm H}$ for any set of flow parameters providing in the transonic accretion solutions. As $T_e$ exceeds the non-relativistic temperature limit $5.8 \times 10^{9}~{\rm K}~(= m_{e}c^{2}/k_{\rm B})$ \cite{novikov-1973} very close to the singularity, the second term in Eq.~(\ref{eq:R-br-emissivity}) starts dominating over the first term. As a result, power density of the electron-electron emission surpasses that of the electron-ion emission. Therefore, one should not neglect the contribution from electron-electron emission to the total emitted power, especially for hot accretion flow (HAF) in AGNs \cite{Yarza-2020} and BHXRBs \cite{rajesh-2010, Dihingia-2017, dihingia-2020}.

\begin{figure*}
	\centering
	\includegraphics[width=\linewidth]{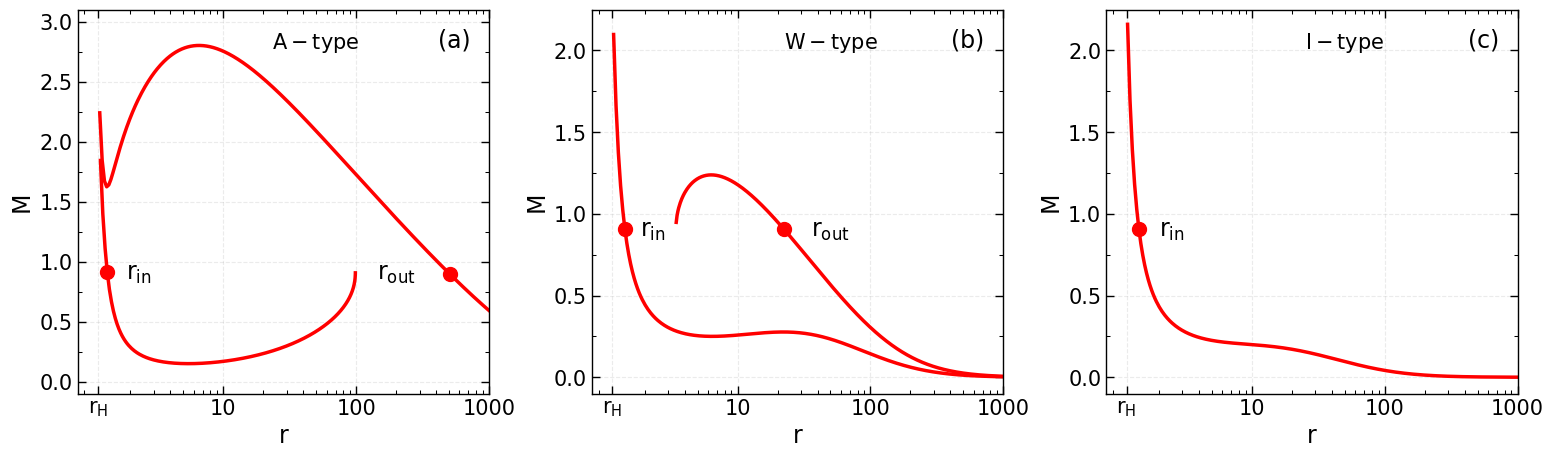}
	\caption{Variation of Mach number ($M$) as a function of radial distance ($r$) for a BH model with $a_{k} = 0.99$. Here, critical points are marked by the filled circles. In this figure, we choose flow parameters as ($\lambda, E$) = ($2.05, 1.0005$), ($2, 1.01$) and ($2, 1.02$) for the respective panels (a-c). See the text for details.}  
	\label{fig:fig3}
\end{figure*}

\begin{figure}
	\centering
	\includegraphics[width=\columnwidth]{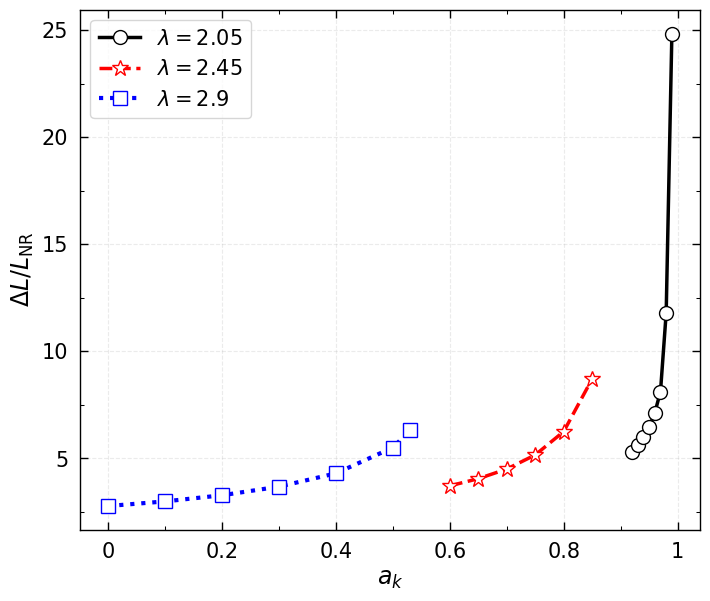}
	\caption{Variation of relative change in the bolometric luminosity $(\Delta L/L_{\rm NR})$ as a function of spin parameter ($a_{k}$) for different angular momentum ($\lambda$). Here, we set $E = 1.0005$. See the text for details.} 
	\label{fig:fig4}
\end{figure}

We then calculate the frequency-integrated disc luminosity for non-relativistic ($L_{\rm NR}$) and relativistic ($L_{\rm R}$) models in Kerr spacetime. Here, we fix the flow energy as $E = 1.0005$, and vary the angular momentum as $\lambda = 2.05$, $2.45$ and $2.9$. We find that the flow possesses multiple critical points ($i.e.$, flows have both inner ($r_{\rm in}$) and outer ($r_{\rm out}$) critical points) for these input parameters. The solutions passing through the outer critical points ($r_{\rm out}$) successfully connect $r_{\rm H}$ to $r_{\rm edge}$ (open solutions\footnote{Open solutions are also called global solutions as they extend from $r_{\rm edge}$ to $r_{0}$. In this paper, all calculations have been carried out for global solutions only.}). However, the solutions passing through the inner critical points ($r_{\rm in}$) unable to connect $r_{\rm H}$ to $r_{\rm edge}$ as they terminate at some $r$ values (closed solutions\footnote{As closed solutions are not physically acceptable, we do not consider such solutions.}). Such flow solution is defined as A-type topology and the example of such accretion solution is presented in Fig.~\ref{fig:fig3}a. The variation of relative change in bolometric luminosity ($\Delta L/L_{\rm NR} = (L_{\rm R} - L_{\rm NR})/L_{\rm NR}$) as a function of spin parameter ($a_{k}$) for different $\lambda$ is presented in Fig.~\ref{fig:fig4}. In this figure, open circles joined by the solid curve (black), open asterisks joined by the dashed curve (red) and open squares joined by the dotted curve (blue) denote the quantity $\Delta L/L_{\rm NR}$ associated with the angular momentum $\lambda = 2.05$, $2.45$ and $2.9$, respectively. It is observed that $\Delta L/L_{\rm NR}$ increases with $a_{k}$, irrespective of $\lambda$ values. It is noteworthy to mention that the change in $\Delta L/L_{\rm NR}$ for R-model is marginal for weakly spinning black holes ($a_{k} \rightarrow 0$), however it becomes significant for rapidly rotating black holes ($a_{k} \rightarrow 1$).

Indeed, the above analysis involving the Kerr spacetime indicates that the relativistic effect ($i.e.$, R-model) becomes important when electron temperature ($T_{e}$) is high and the central source spins rapidly. Hence, it seems reasonable to consider the R-model in carrying out the emission analyses for wide ranges of $T_{e}$ and $a_{k}$. Accordingly, we study the emission properties of the accretion flow considering R-model in the subsequent sections.

\section{Luminosity distribution for Kerr black hole}
\label{sec:SED-Kerr-BH}

\begin{figure}
	\centering
	\includegraphics[width=0.9\columnwidth]{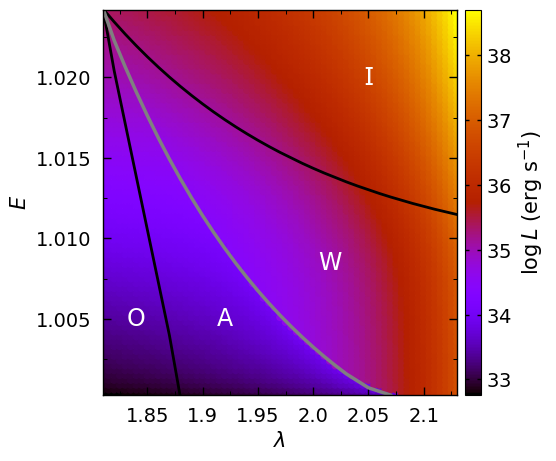}
	\caption{Division of parameter space in $\lambda - E$ plane according to the nature of flow solutions in the BH model. Four regions are marked as O, A, W and I. Here, we choose $a_{k} = 0.99$. Color map denotes the 2D projection of 3D plot of $\lambda$, $E$ and $L$.  Vertical color bar indicates $L$ in $\rm erg~s^{-1}$. See the text for details.}  
	\label{fig:fig5}
\end{figure}

In this section, we study the disc luminosity corresponding to different flow solutions in Kerr spacetime, where $a_{k} = 0.99$ is chosen for the purpose of representation. We calculate the parameter space according to the flow topologies (O, A, W and I-types) in the angular momentum ($\lambda$) and energy ($E$) plane and depict it in Fig.~\ref{fig:fig5}. These flow topologies have different physical properties. For example, O-type solutions only contain outer critical points ($r_{\rm out}$), whereas I-type solutions only have inner critical points ($r_{\rm in}$). In both cases, flow solutions passing through the critical points smoothly connect $r_{\rm edge}$ and $r_{\rm H}$ ($i.e.,$ open solutions (see Figs.~\ref{fig:fig2}a and \ref{fig:fig3}c for O and I-type solutions, respectively)). Therefore, the flow topologies in which the existing accretion solutions pass through single critical points $r_{\rm out}$ or $r_{\rm in}$ can be distinguished as O or I-type flow topologies. However, for both the A and W-type flow topologies the accretion solutions must contain multiple critical points ($i.e.,$ $r_{\rm in}$ and $r_{\rm out}$). In the A-type topology, solution passes through $r_{\rm in}$ ($r_{\rm out}$) remain closed (open) (see Fig.~\ref{fig:fig3}a) and satisfy the condition $\dot{\mathcal M}_{\rm in} > \dot{\mathcal M}_{\rm out}$, where $\dot{\mathcal M}_{\rm in}$ and $\dot{\mathcal M}_{\rm out}$ are the entropy accretion rate at $r_{\rm in}$ and $r_{\rm out}$, respectively. On the contrary, for the W-type flow topology, solution passes through $r_{\rm in}$ ($r_{\rm out}$) remain open (closed) (see Fig.~\ref{fig:fig3}b) and satisfies the condition $\dot{\mathcal M}_{\rm in} < \dot{\mathcal M}_{\rm out}$. Therefore, it is useful to separate the effective domain of the parameter space for different flow solutions ($e.g.$ O, A, W and I) in terms of the entropy accretion rate along with the number of critical points \cite[see][and references therein]{dihingia-2018, dihingia-2020a, Dihingia-2019, Patra-2022, Sen-2022}. In the figure, the region within black curves provides flow solutions containing multiple critical points (both $r_{\rm in}$ and $r_{\rm out}$). The gray curve separates the regions A and W and it corresponds to $\dot{\mathcal M}_{\rm in} = \dot{\mathcal M}_{\rm out}$. We make use of the global accretion solutions corresponding to a given set of $(\lambda, E)$ and calculate the frequency-integrated luminosity $(L)$ using Eq.~(\ref{eq:bolometric-L}). The obtained results are presented using color map in Fig. \ref{fig:fig5}, where the color bar at the right refers the range of $L$ in $\rm erg~s^{-1}$. Note that for a given $\lambda$, $L$ increases with $E$. Similarly, for a given $E$, $L$ increases with $\lambda$. This happens because of the fact that as $\lambda$ (or $E$) is increased keeping $E$ (or $\lambda$) fixed, the flow temperature increases due to the shifting of critical point towards the horizon (see Table 2 in \cite{Patra-2022}). In general, the critical points for I-type solutions are formed very close to the central singularity that results higher the luminosity values compared to the other flow orbits. We further investigate the spectral features of different global accretion solutions and delineate them in the subsequent sections.

\begin{figure*}
	\centering
	\includegraphics[width=\linewidth]{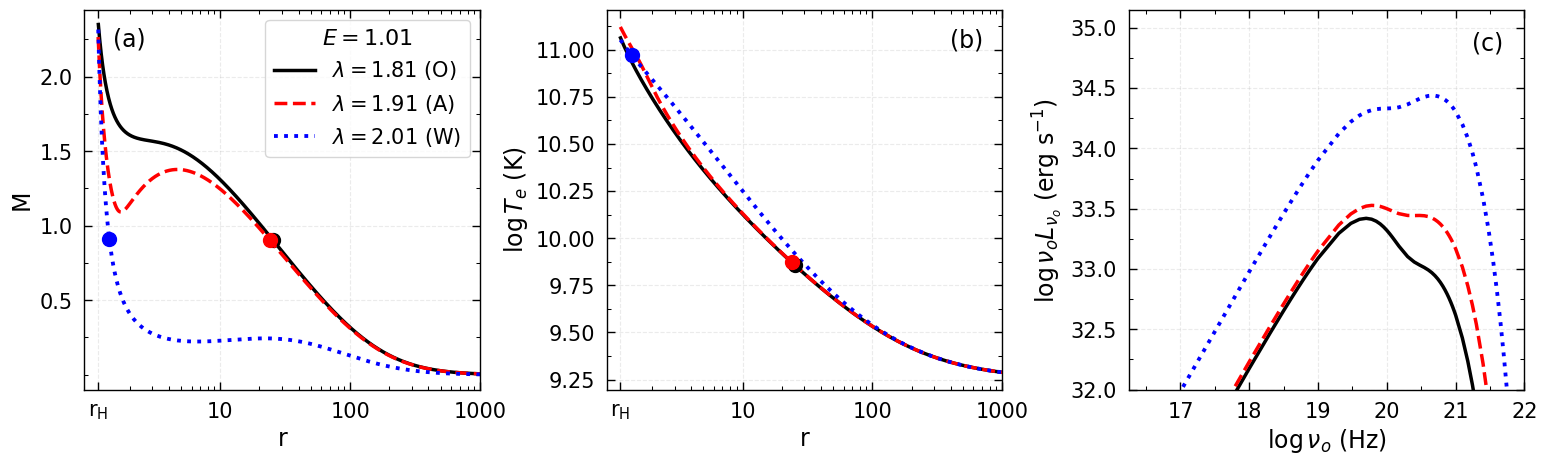}
	\caption{Global accretion solutions ($M$ vs. $r$ plots in panel (a)), electron temperature profiles ($T_{e}$ vs. $r$ plots in panel (b)) and spectral energy distributions ($\nu_{o}L_{\nu_o}$ vs. $\nu_o$ in panel (c)) for O, A and W-type flow topologies. Angular momentum ($\lambda$) for different flow solutions have been marked in panel (a). Here, critical points are marked by the filled circles. In this figure, we choose a BH model with $a_{k} = 0.99$. And, flow energy is taken as $E = 1.01$. See the text for details.}  
	\label{fig:fig6}
\end{figure*}

\begin{table}
	\centering
	\caption{Flow energy $(E)$, angular momentum $(\lambda)$, inner critical points $(r_{\rm in})$, outer critical points $(r_{\rm out})$ and type of accretion solutions are presented in columns 1-5. These solutions are associated with the SEDs in Figs.~\ref{fig:fig6} and~\ref{fig:fig7}.}
	\label{tab:table-1}
	\begin{ruledtabular}
		\begin{tabular}{lcccc}
			$E$ & $\lambda$ & $r_{\rm in}$ & $r_{\rm out}$ & Type\\
			\hline
			1.010 & 1.81 & --- & 25.3308 & O\\
			& 1.91 & 1.7184 & 23.9365 & A\\
			& 2.01 & 1.3981 & 22.2109 & W\\
			1.015 & 1.81 & --- & 15.7051 & O\\
			& 1.87 & 1.9952 & 14.6330 & A\\
			& 1.93 & 1.6078 & 13.1974 & W\\
			& 1.99 & 1.4136 & --- & I\\
			1.020 & 1.81 & --- & 10.3324 & O\\
			& 1.83 & 2.5479 & 9.7945 & A\\
			& 1.85 & 2.1608 & 9.0972 & W\\
			& 1.87 & 1.9008 & --- & I\\
		\end{tabular}
	\end{ruledtabular}
\end{table}

Next, we explore the spectral energy distributions (SEDs) corresponding to the four accretion solution topologies ($i, e.$, O, A, W and I-types) in the Kerr spacetime. Here, we choose the flow energy as $E = 1.01$, and vary the angular momentum of the flow as $\lambda = 1.81, 1.91$ and $2.01$ to obtain the O, A and W-type solutions (see Fig.~\ref{fig:fig5}). We identify the global accretion solutions that pass through either outer or inner critical points as $r_{\rm out} = 25.3308$ (O-type solution), $r_{\rm out} = 23.9365$ (A-type solution), and $r_{\rm in} = 1.3981$ (W-type solution), respectively (see Table~\ref{tab:table-1} for details). These solutions are shown in Fig.~\ref{fig:fig6}a, where solid (black), dashed (red) and dotted (blue) curves are for O, A and W-type solutions. In Fig.~\ref{fig:fig6}b, we show the profile of electron temperature ($T_{e}$) corresponding to the solutions presented in Fig.~\ref{fig:fig6}a. Using Eq.~(\ref{eq:monochromatic-L}), we compute the SED for these solutions and present the obtained results in Fig.~\ref{fig:fig6}c, where the variation of $\nu_o L_{\nu_o}$ is plotted as a function of frequency $\nu_o$. We observe that the bremsstrahlung emission dominates at $\nu_o \approx 10^{20}~{\rm Hz}$ \cite{Yarza-2020} and has a sharp cut-off at $\nu_o \approx 10^{22}~{\rm Hz}$ ($= k_{\rm B}T_{e0}/h$)~\cite{novikov-1973}, where $T_{e0} \approx 10^{11}~{\rm K}$ is the electron temperature at disc inner edge ($r_{0}$). Note that SEDs for O and A-type solutions are almost identical at the low frequencies, however a reasonable difference is seen at the high frequencies. For A-type solution, the critical point is formed at relatively smaller distance compared to the O-type solution that results the higher temperature at the inner part of the disk (see Fig.~\ref{fig:fig6}b). Consequently, the A-type solutions generate high luminous power spectra than the O-type solutions. In comparison, SED for W-type solution is extremely high compared to the O and A-type solutions at the both frequency ends. Since the critical point for W-type solution is formed near the central object, the overall temperature profile is enhanced and therefore, hot plasma in the accretion disc produces high luminous spectra compared to O and A-type solutions.

\begin{figure*}
	\centering
	\includegraphics[width=\linewidth]{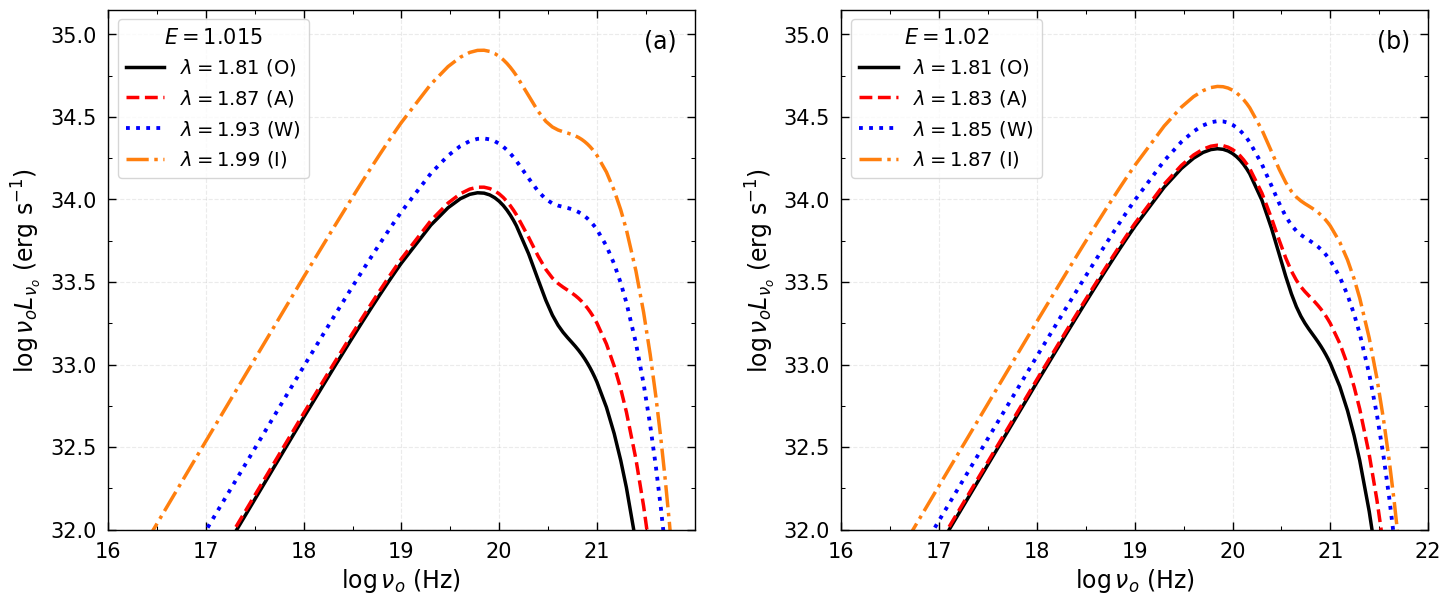}
	\caption{Spectral energy distributions (SEDs) for O, A, W and I-type solutions corresponding to the flow energies at $E = 1.015$ (panel a) and $1.02$ (panel b). Angular momentum ($\lambda$) for different flow solutions are marked in each panels. Here, we consider a BH model with $a_{k} = 0.99$. See the text for details.}  
	\label{fig:fig7}
\end{figure*}

\begin{figure*}
	\centering
	\includegraphics[width=\linewidth]{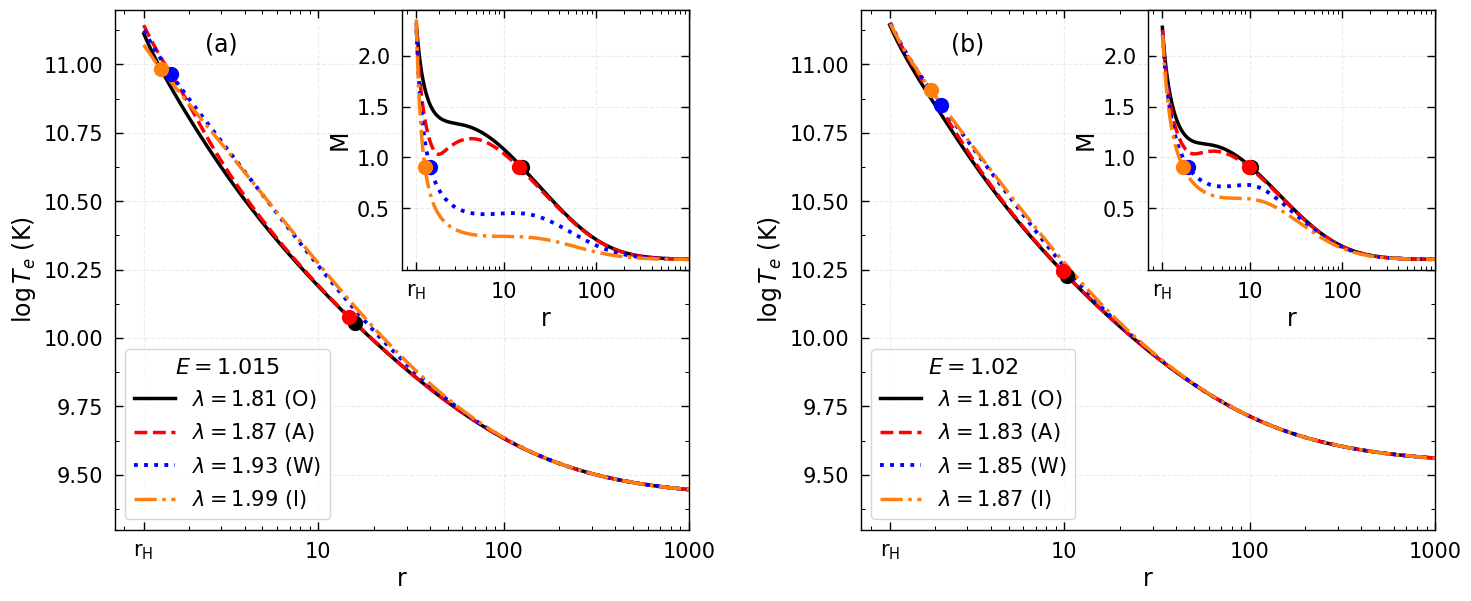}
	\caption{Variation of electron temperature ($T_{e}$) as a function radial distance ($r$) for Fig.~\ref{fig:fig7}. Inset panels denote the respective global accretion solutions ($M$ vs. $r$ plots). In all cases, critical points are marked by the filled circles. See the text for details.}  
	\label{fig:fig8}
\end{figure*}

We continue the study of SEDs for higher flow energies and in Figs.~\ref{fig:fig7}a-b, we present the SEDs corresponding to O, A, W and I-type flow solutions that are obtained for $E = 1.015$ and $1.02$, respectively. In both cases, we find that I-type solutions result maximum luminous power spectra compared to the other flow solutions. The details of the critical point locations for these accretion solutions and their input flow parameters are tabulated in Table~\ref{tab:table-1}. As the critical points of the I-type solutions are formed at relatively smaller distances (close to the singularity) than O, A and W-type solutions, we get higher SEDs for I-type solutions. For the purpose of clarity, we present the electron temperature $(T_{e})$ profiles in Figs.~\ref{fig:fig8}a-b corresponding to the flow solutions given in Figs.~\ref{fig:fig7}a-b. It is noticed that the temperature distributions in the disc increased when the critical points are drifted toward the singularity. Therefore, the flows that pass through the inner critical points $(r_{\rm in})$ possess high electron temperature profiles than those through the outer critical points $(r_{\rm out})$. Needless to mention that the obtain SEDs for these four types of accretion solutions strongly depend on the input flow parameters, namely energy ($E$) and angular momentum ($\lambda$).

\section{Spectral analysis in deformed spacetime}  
\label{sec:deformed-spacetime}

In Section~\ref{sec:bremsstrahlung}, we investigate the usefulness of the relativistic model (R-model) over the non-relativistic model (NR-model) for hot accretion flow (HAF) in the usual Kerr spacetime. In this section, we examine how the deformation of spacetime affects the bolometric disc luminosity as well as the spectral energy distributions in both models ($i. e.$, NR-model and R-model).

\begin{figure}
	\centering
	\includegraphics[width=\columnwidth]{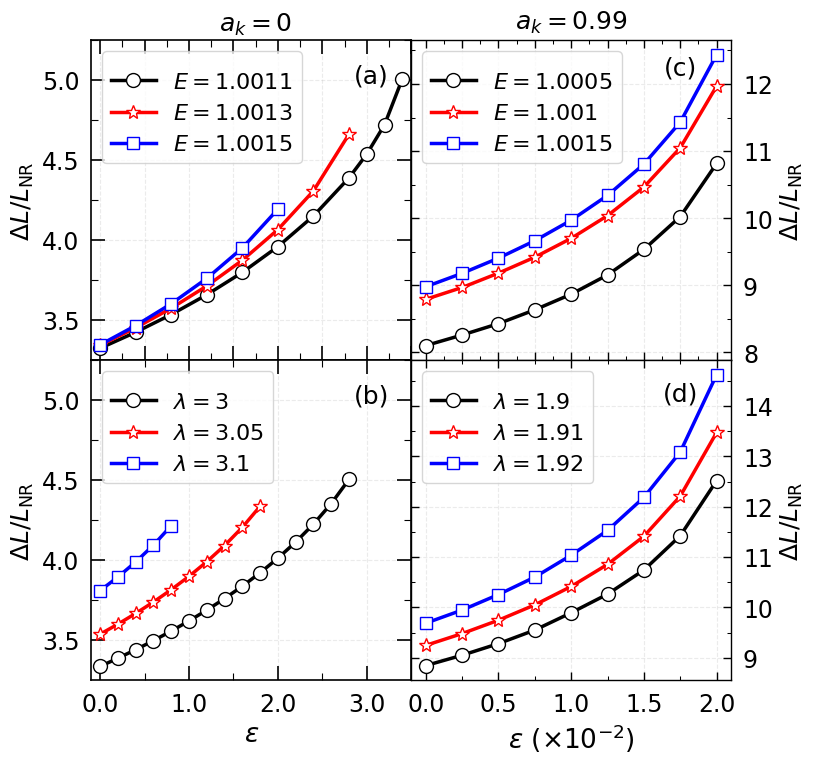}
	\caption{Variation of relative change in bolometric luminosity ($\Delta L/L_{\rm NR}$) as a function of deformation parameter ($\varepsilon$) for $a_{k} = 0$ (left panels) and for $a_{k} = 0.99$ (right panels). The flow angular momentum ($\lambda$) and energy ($E$) are marked in each panel. See the text for details.}
	\label{fig:fig9}
\end{figure}

\subsection{Effect of $\varepsilon$ on disc luminosity for NR and R-models}
\label{sec:Comparison-R-NR-model}

In order to understand the effect of deformation parameter $(\varepsilon)$ on the disc luminosity, we examine the relative change of bolometric disc luminosity ($\Delta L/L_{\rm NR}$) considering both weakly rotating ($a_{\rm k}\rightarrow 0$) and rapidly rotating ($a_{\rm k}\rightarrow 1$) black holes and depict the obtained results in Fig.~\ref{fig:fig9}. In the left panels, we show the variation of $\Delta L/L_{\rm NR}$ as a function of $\varepsilon$ for $a_{k} = 0$. In panel (a), we choose $\lambda = 3$ and vary energy ($E$). Open circles (black), open asterisks (red) and open squares (blue) joined with solid lines denote the results obtained for flow energies $E = 1.0011$, $1.0013$ and $1.0015$, respectively. We observe that for a given $E$ value, the relative increment in $L_{\rm R}$ with respect to $L_{\rm NR}$ increases with $\varepsilon$. We also notice that for a fixed $\varepsilon$, $\Delta L/L_{\rm NR}$ increases with $E$. In panel (b), we plot $\Delta L/L_{\rm NR}$ as a function of $\varepsilon$ for fixed $E = 1.0012$, where $\lambda$ is varied. Open circles (black), open asterisks (red) and open squares (blue) joined with solid lines denote the results obtained for $\lambda = 3, 3.05$ and $3.1$, respectively. We find that $\Delta L/L_{\rm NR}$ increases with $\varepsilon$, irrespective of $\lambda$ values. And, for a fixed $\varepsilon$, $\Delta L/L_{\rm NR}$ increases with $\lambda$. Similarly, in the right panels, we present the variation of $\Delta L/L_{\rm NR}$ as a function of $\varepsilon$ for $a_{k} = 0.99$. In panel (c), we fix $\lambda = 1.9$, and vary energy as $E = 1.0005$, $1.001$ and $1.0015$. The obtained results are plotted using open circles (black), open asterisks (red) and open squares (blue) joined with solid lines. Finally, in panel (d), we set $E = 1.002$ and vary angular momentum as $\lambda = 1.9$, $1.91$ and $1.92$ and the obtained results are shown using open circles (black), open asterisks (red) and open squares (blue) joined with solid lines. As in panel (a) and (b), we observe similar variation in (c) and (d), however $\Delta L/L_{\rm NR}$ attains higher values for $a_{k} = 0.99$ compared to the results for $a_{k} = 0$, which is consistent with the results presented in Fig.~\ref{fig:fig4}. With this, we indicate that while exploring spectral analysis in non-Kerr spacetime, the effect of bremsstrahlung emission in R-model seems very much relevant compared to the NR-model.

\subsection{Effect of $\varepsilon$ on SEDs}
\label{sec:SED-deformed-BH}

\begin{figure*}
	\centering
	\includegraphics[width=\textwidth]{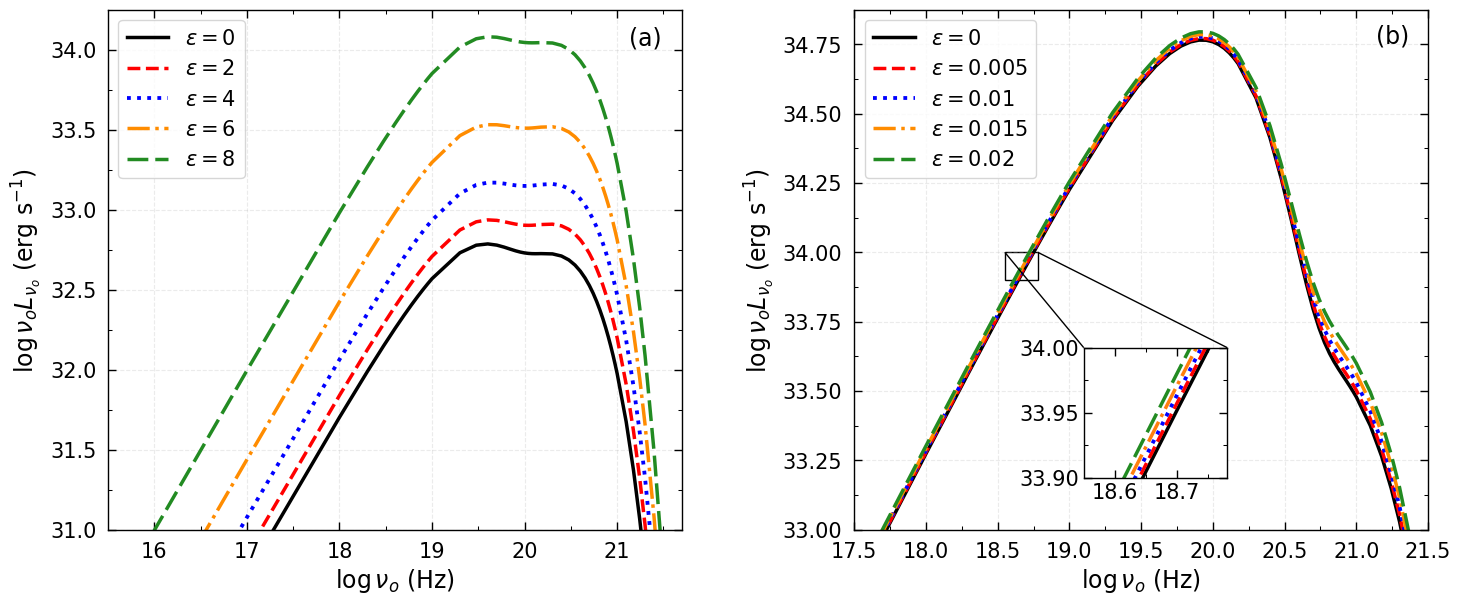}
	\caption{Effect of deformation parameter ($\varepsilon$) on the spectral energy distributions. In panel (a), results are presented for $a_{\rm k} = 0$, $E=1.005$ and $\lambda = 2.9$, whereas the same is shown in panel (b) for $a_{\rm k} = 0.99$, $E=1.0215$ and $\lambda = 1.85$. See the text for details.}  
	\label{fig:fig10}
\end{figure*}

In this section, we examine how the deformation of spacetime affects the spectral energy distributions (SEDs). While doing so, we consider two black hole models: one with weakly rotating ($a_{k} \rightarrow 0$) central source and the other with rapidly rotating ($a_{k} \rightarrow 1$) central source. For $a_{k} = 0$, we choose $E = 1.005$ and $\lambda = 2.9$, and calculate the SEDs for a set of deformation parameters. To evaluate the thermal bremsstrahlung spectrum, we use the same methodology as described in Section~\ref{sec:SED-Kerr-BH}. The obtained results are presented in Fig.~\ref{fig:fig10}a, where solid (black), dashed (red), dotted (blue), dot-dashed (orange) and long-dashed (green) curves represent SEDs corresponding to the deformation parameter $\varepsilon = 0$, $2$, $4$, $6$ and $8$, respectively. It is observed that SEDs increase significantly at both low and high-frequency ends when $\varepsilon$ increases. We find that the above set of input model parameters yields only I-type accretion solutions with $r_{\rm in} = 6.7663, 5.5027, 4.5666, 3.9224$ and $3.5186$ for increasing $\varepsilon$ values starting from $0$ to $8$. Next, we fix $a_{k} = 0.99$, $E= 1.0215$ and $\lambda = 1.85$, and calculate the SEDs for a set of $\varepsilon$ values. We present the obtained results in Fig.~\ref{fig:fig10}b, where solid (black), dashed (red), dotted (blue), dot-dashed (orange) and long-dashed (green) curves are for $\varepsilon = 0$, $0.005$, $0.01$, $0.015$ and $0.02$, respectively. Note that the allowed range of $\varepsilon$ yielding the black hole solutions generally decreases with $a_{\rm k}$ \cite[see Fig.~17 of][]{Patra-2022} and hence, we use relatively smaller value of $\varepsilon$ for $a_{\rm k}=0.99$ compared to the same used for weakly rotating black hole. As before, these SEDs are calculated using I-type global accretion solutions obtained for above set of $\varepsilon$ and they become transonic after crossing the inner critical points at $r_{\rm in} = 2.1343, 2.082, 2.0261, 1.9655$ and $1.898$, respectively. We note that SEDs are shifted at higher values as $\varepsilon$ is increased, however their differences are barely noticeable as the range of $\varepsilon$ is small.       

\subsection{Luminosity distribution for naked singularity}
\label{sec:SED-NS}  

\begin{figure*}
	\centering
	\includegraphics[width=\linewidth]{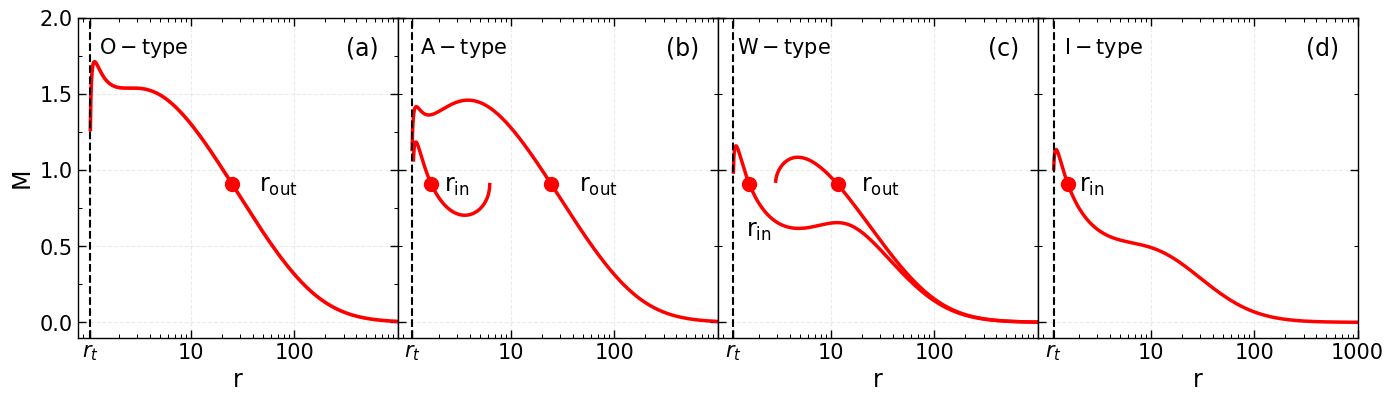}
	\caption{Variation of Mach number ($M$) as a function of radial distance ($r$) for NS model. Here, we choose $a_{k} = 0.99$ and $\varepsilon = 0.03$. Results presented in panels (a-d) are obtained for ($\lambda, E$) = ($1.82, 1.01$), ($1.86, 1.01$), ($1.86, 1.0175$) and ($1.86, 1.025$), respectively. In each panel, filled circles denote the critical points ($r_{\rm in}$ and $r_{\rm out}$) and dotted vertical line indicate the inner edge  ($r_{0} = r_{t}$) of the disc. See the text for details.}  
	\label{fig:fig11}
\end{figure*}

\begin{figure}
	\centering
	\includegraphics[width=0.9\columnwidth]{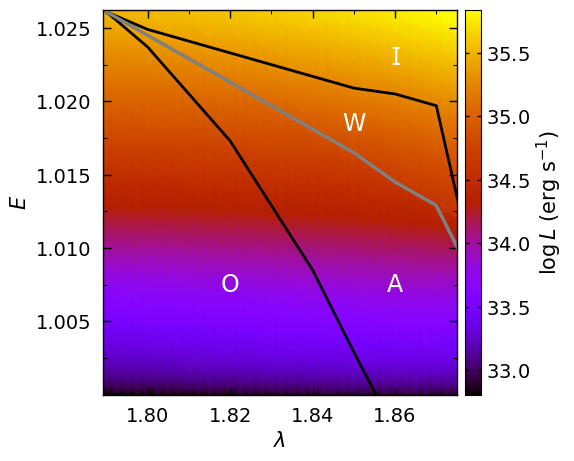}
	\caption{Division of parameter space in $\lambda - E$ plane according to the behavior of flow solutions (O, A, W and I-types) in the NS model. Color map denotes 2D projection of 3D plot of $\lambda$, $E$ and $L$. Vertical color bar at the right denotes $L$ in ${\rm erg~s^{-1}}$. Here, we choose $a_{\rm k}=0.99$ and $\varepsilon = 0.03$. See the text for details.}  
	\label{fig:fig12}
\end{figure}  

In this section, we examine disc luminosity ($L$) variation of accretion flow around the naked singularity (NS). In doing so, we consider a rapidly rotating central object of spin $a_{k} = 0.99$ with deformation parameter $\varepsilon = 0.03$. This choice of parameters provides accretion solution in NS as Eq. (\ref{eq:event-horizon}) does not yield real root of $r_{\rm H}$. With this, we calculate the accretion solutions around naked singularity by freely varying flow energy ($E$) and angular momentum ($\lambda$), and separate the parameter space in $\lambda - E$ plane according to the nature of accretion solution topologies ($i.e.$, O, A, W and I-type) around naked singularity. In Fig.~\ref{fig:fig11}, we show the nature of O, A, W and I-type accretion solutions ($i.e.$, $M = v/C_{s}$ vs. $r$ plots) for NS model. The detailed physical properties of various accretion flow solutions in the naked singular spacetime are examined in~\cite{dihingia-2020a, Patra-2022}. Here, the obtained results are presented in Fig.~\ref{fig:fig12}, where the identified regions are marked. The region bounded by the black curves yield multiple critical points which is further sub-divided according to the entropy condition where gray curve corresponds to $\dot{\mathcal M}_{\rm in} = \dot{\mathcal M}_{\rm out}$. We further calculate the frequency-integrated disc luminosity $(L)$ utilizing the accretion solutions from the entire parameter space and depict the obtained results using colors in Fig.~\ref{fig:fig12}. To obtain $L$, we integrate Eq.~(\ref{eq:bolometric-L}) from the disk inner edge $r_{0}$ to the outer edge $r_{\rm edge} = 1000$. It may be noted that for NS model, flows co-rotate along a surface (often called naked surface) very close to the singularity (see Fig.~\ref{fig:fig11}). Hence, the flow is truncated at a radius $r_{t}$ (see Fig.~\ref{fig:fig11}) near the singularity and we consider $r_{t}$ as the inner edge of the disc for NS model ($i.e., r_{0} \sim  r_{t}$). In the figure, vertical color bar at the right denotes the range of $L$. We find that for a given $\lambda$ (or $E$), $L$ increases as $E$ (or $\lambda$) is increased. Therefore, it is evident that I-type solutions generally yield larger disc luminosity $L$ compared to the other accretion solutions around NS.

\begin{figure*}
 	\centering
 	\includegraphics[width=\linewidth]{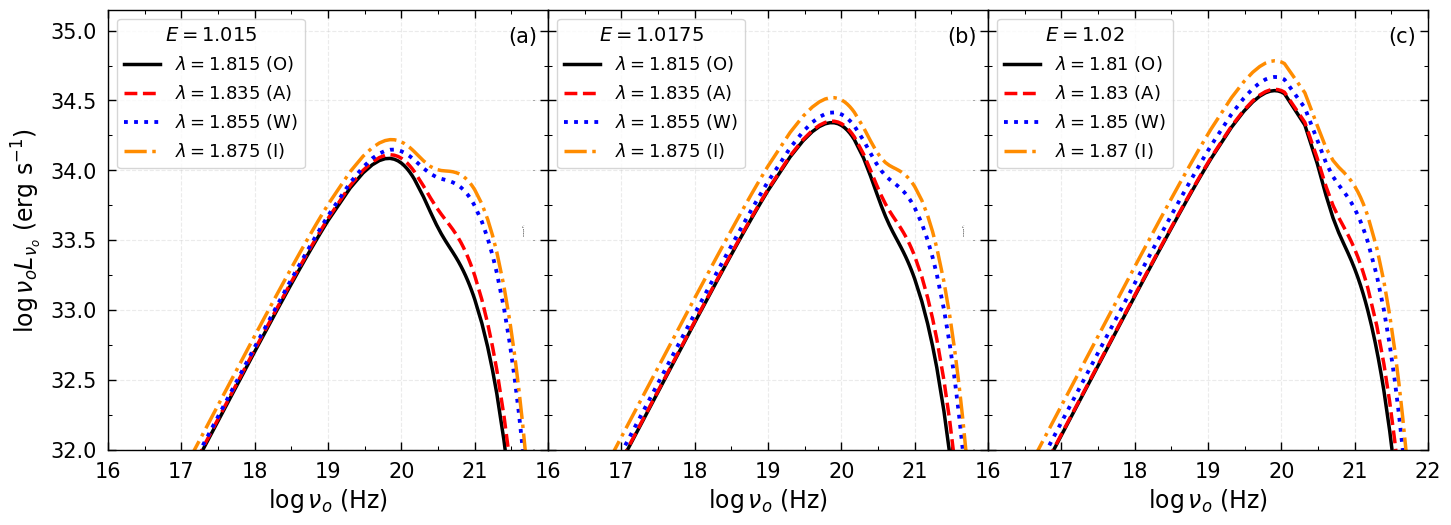}
 	\caption{Spectral energy distributions (SEDs) corresponding to different accretion solutions (O, A, W and I-types) around NS for flow energies $E = 1.015$ (panel a), $1.0175$ (panel b) and $1.02$ (panel c). In each panel, solid (black), dashed (red), dotted (blue) and dot-dashed (orange) corves denote results for different $\lambda$ which are marked. Here, we choose $a_{\rm k}=0.99$ and $\varepsilon = 0.03$. See the text for details.}  
 	\label{fig:fig13}
\end{figure*}

\begin{table}
	\centering
	\caption{Flow energy $(E)$, angular momentum $(\lambda)$, inner critical points $(r_{\rm in})$, outer critical points $(r_{\rm out})$, type of accretion solutions in NS model are tabulated in columns1-5. These solutions are used to obtain the SEDs shown in Fig.~\ref{fig:fig13}.}
	\label{tab:table-2}
	\begin{ruledtabular}
		\begin{tabular}{lcccc}
			$E$ & $\lambda$ & $r_{\rm in}$ & $r_{\rm out}$ & Type\\
			\hline
			1.015 & 1.815 & --- & 15.6082 & O\\
			& 1.835 & 2.1152 & 15.2698 & A\\
			& 1.855 & 1.6405 & 14.8097 & W\\
			& 1.875 & 1.4214 & --- & I\\
			1.0175 & 1.815 & --- & 12.6171 & O\\
			& 1.835 & 2.0377 & 12.2021 & A\\
			& 1.855 & 1.6895 & 11.7311 & W\\
			& 1.875 & 1.4057 & --- & I\\
			1.02 & 1.81 & --- & 10.2935 & O\\
			& 1.83 & 2.0806 & 9.7440 & A\\
			& 1.85 & 1.7361 & 7.4222 & W\\
			& 1.87 & 1.4403 & --- & I\\
			
		\end{tabular}
	\end{ruledtabular}
\end{table}                              
   
Next, we examine the spectral energy distributions (SEDs) for different accretion solutions in NS model. Here, we choose $E = 1.015, 1.0175$ and $1.02$, and vary $\lambda$ to calculate SEDs for O, A, W and I-type accretion solutions. The obtained results are shown in Fig.~\ref{fig:fig13}a-c, where solid (black), dashed (red), dotted (blue) and dot-dashed (orange) curves represent SEDs associated with O, A, W and I-type flow solutions, respectively. It is clear from the figures that obtained SEDs alter due to the variation of flow parameters ($i.e.$, $E$ and $\lambda$). We further observe that I-type flow solutions concede higher SEDs over the other flow solutions as indicated in Fig.~\ref{fig:fig12}. The chosen parameters to obtain the above figures are tabulated in Table~\ref{tab:table-2}.

\subsection{Comparison between SED$\small{\text{s}}$ for BH and NS models}
\label{sec:SED-BH-NS}              

\begin{figure*}
	\centering
	\includegraphics[width=\linewidth]{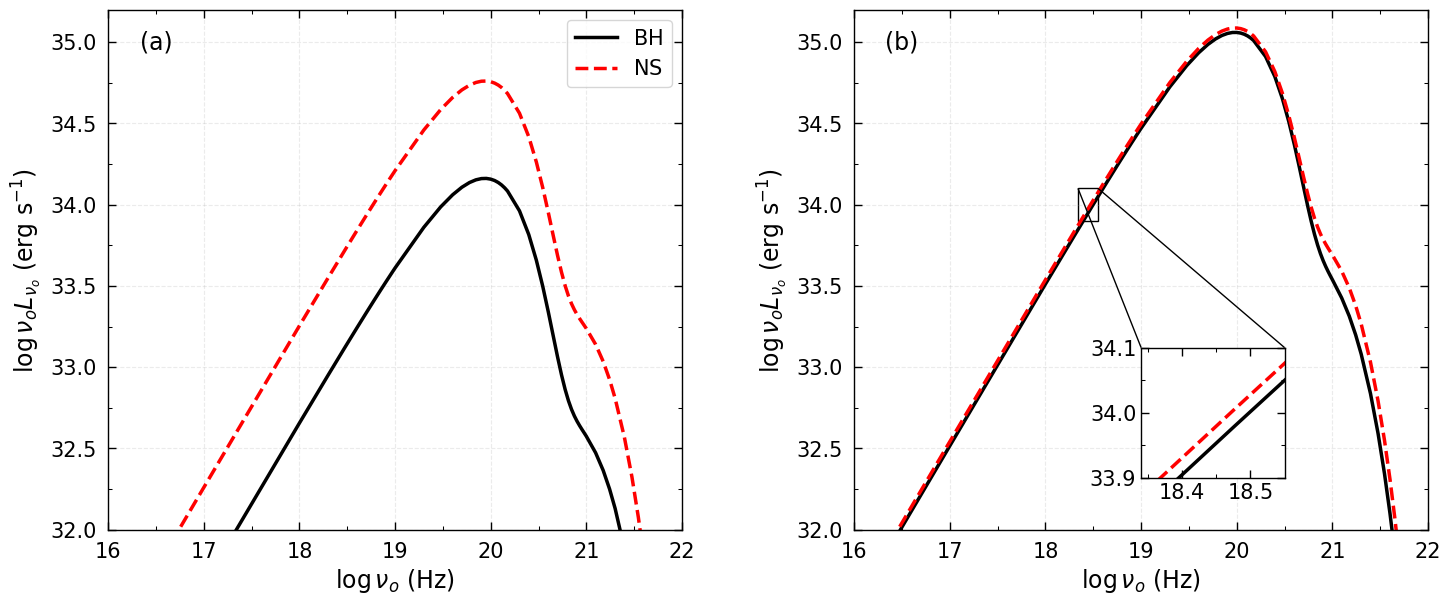}
	\caption{Comparison of SEDs obtaind from BH and NS models. In panel (a), we set $a_{k} = 0.5$, $E = 1.0225$ and $\lambda =1.7$, respectively. Solid (black) and dashed (red) curves denote results for BH ($\varepsilon = 6$) and NS ($\varepsilon = 7$). In panel (b), we choose $a_{k} = 0.99$, $E = 1.025$ and $\lambda =1.83$, and SED for BH and NS are obtained for $\varepsilon = 0.02$ and $0.04$, respectively. See the text for details.}  
	\label{fig:fig14}
\end{figure*}

In this section, we compare the SEDs obtained from BH and NS models. While doing this, we choose the model parameters as $E = 1.0225$, $\lambda =1.7$ and $a_{k} = 0.5$ and calculate the accretion solutions for BH and NS models using $\varepsilon = 6$ and $7$, respectively. This choice of model parameters results I-types accretion solutions with $r_{\rm in} = 2.5067$ for BH model and $r_{\rm in} = 1.9059$ for NS model and the solutions extend up to the outer edge $r_{\rm edge} = 1000$. We calculate the SEDs for these solutions and obtained results are depicted in Fig.~\ref{fig:fig14}a, where solid (black) and dashed (red) are for BH and NS models. It is evident that SED for NS model is significantly higher at both low and high frequencies ends compared to the same for BH model. In Fig.~\ref{fig:fig14}b, we display the results for $a_{k} = 0.99$, where flow parameters are chosen as $E = 1.025$ and $\lambda =1.83$, and $\varepsilon = 0.02$ and $0.04$ render I-type accretion solutions for BH and NS models, respectively. Here, $r_{\rm in} = 2.1193$ for BH model and $r_{\rm in} = 1.737$ for NS model. As before, SED is more for NS model compared to BH model although the difference seems marginal in all frequency range. This findings are in agreement with \citet{Tahelyani-2022}.

\section{Summary and Discussion}
\label{sec:Summary}  

In this paper, we investigate thermal bremsstrahlung emission from the accretion disc in black hole and naked singular spacetime. We consider the Johannsen-Psaltis (JP) non-Kerr spacetime, which describes the central object as a black hole or naked singularity depending on the choice of spin ($a_{k}$) and deformation parameters ($\varepsilon$). In hot accretion flow (HAF), the relativistic effect and electron-electron emission modify the non-relativistic bremsstrahlung emissivity. We use the Novikov-Thorne equation for emission coefficient \cite{novikov-1973}, which includes both the effects mentioned above in addition to the electron-ion emission. A basic structure of general relativistic accretion flow around a stationary and axisymmetric compact object has been developed. We obtain the expressions of monochromatic and bolometric disc luminosity in an observer frame at spatial infinity. We summarize our findings below.

\begin{itemize}
	
	\item We find that for hot accretion flow (HAF), bremsstrahlung emissivity in the relativistic model (R-model) is larger than the non-relativistic model (NR-model). Moreover, the electron-electron emission near the singular point significantly enhances the emissivity in R-model compared to NR model. Accordingly, we compute the relative change in disc luminosity ($\Delta L/L_{\rm NR}$) for R-model with respect to NR-model. The obtained results clearly indicate that the relativistic effect becomes dominant for black holes with higher spin values. Hence, we emphasize to consider the bremsstrahlung emission in studying the emission properties of the accretion flow using R-model.
	
	\item For black hole model, we calculate the disc luminosity ($L$) by freely varying flow energy ($E$) and angular momentum ($\lambda$). We find that for a given $E$, $L$ increases with $\lambda$. Likewise, for a particular $\lambda$, $L$ increases with $E$. Furthermore, we calculate the spectral energy distributions (SEDs) associated with the different flow topologies (O, A, W and I-types) in $\lambda - E$ parameter space. We notice that SEDs strongly depend on the flow parameters ($\lambda, E$). At the low frequency region, SEDs for O and A-type solutions differ very small compared to the results in high frequency region. But, SEDs for W and I-type solutions significantly differ from O and A-type topologies at both lower and higher frequencies, especially for low energy accretion flows. We also observe that the luminosity distribution for I-type solutions are higher compared to other flow solutions.
	
	\item We examine the role of deformation parameter ($\varepsilon$) in controlling the disc luminosity ($L$) when relativistic effect is invoked. We find that $\Delta L/L_{\rm NR}$ increases with $\varepsilon$, irrespective of the flow parameters ($\lambda, E$). This clearly indicates the importance of relativistic effect in studying the emission properties of accretion flow in deformed spacetime. We also explore the effect of $\varepsilon$ on the disc luminosity spectrum and find that SEDs increase with $\varepsilon$. Therefore, the accretion disc around a non-Kerr black hole is more luminous than the usual Kerr black hole. 
	
	\item In the naked singularity model, we subdivide a parameter space in $\lambda - E$ plane according to the nature of flow solutions. In these regions, we compute $L$ values, where we observe that it increases with $\lambda$ and $E$. Further, we obtain luminosity distributions for different flow topologies in $\lambda - E$ parameter space. We find that SEDs for O and A-type solutions are almost identical at both low and high frequency ends. On the contrary, SEDs for W and I-type solutions moderately differ form O and A-type solutions for all flow energies. This result differs from the black hole model, where we see that SEDs deviate too much in a given frequency range. However, like black hole model, we obtain maximum luminous power spectrum for the I-type solutions compared to the other flow solutions.
	
	\item From a comparative study between the SEDs, we infer that luminosity distributions for naked singular spacetime are higher than the black hole, which agrees with~\citet{Tahelyani-2022}.   
	     
\end{itemize}

With the above findings, it is useful to indicate the observation implications of the present model formalism. In reality, the wide band spectral modelling of the observed spectrum can decipher the physical properties of the black hole, namely mass, spin and accretion rate \cite[and references therein]{Chakrabarti-1995,gou-2009,Nandi-etal2012,Iyer-etal2015,nandi-2018, das-2021, mondal-2022a, heiland-2023, sreehari-2020,majumder-2022}. Accordingly, for the first time to the best of our knowledge, we make an effort to calculate the model spectrum considering the thermal bremsstrahlung emissions from the accretion disk in the realm of deformed spacetime. Indeed, the obtained model spectra are quantitatively different for BH and NS models, however, their typical natures appears to be qualitatively resemblant. This possibly happens because of the fact that the present formalism is developed only involving the thermal bremsstrahlung emissions neglecting the other relevant radiative processes to avoid complexity.

Finally, we mention the limitation of this work. The magnetic fields are ubiquitous in the accretion disk around gravitating objects. Accordingly, in presence of magnetic fields, the fluid momentum equation is altered due to magnetic pressure and induction equation is introduced for the evolution of magnetic fields along the streamline \cite[see][and references therein]{Gammie-etal2003,Mckinney-Gammie2004,mitra-2022}. For simplicity, we do not consider the magnetic fields in the accretion disc and hence, we refrain from using synchrotron emissions.  We also do not incorporate Comptonization. Further, we neglect the effect of viscosity in the disc. Needless to mention that all these physical processes are relevant in the context of accretion flow dynamics. However, implementation of these processes is beyond the scope of the present paper and we plan to take these up in future study.

\section*{Data availability statement}

The data underlying this article will be available with reasonable request.

\section*{Acknowledgment}

Authors thank the anonymous referee for valuable comments to improve the quality of the paper. The work of SP is supported by the University Grants Commission (UGC), Government of India, under the scheme Senior Research Fellowship (SRF). BRM is supported by a START-UP RESEARCH GRANT (No. SG/PHY/P/BRM/01) from the Indian Institute of Technology Guwahati (IITG), India. The work of SD is supported by the Science and Engineering Research Board (SERB), India, through grant MTR/2020/000331.    
\bibliographystyle{apsrev}
\bibliography{references}

\end{document}